\documentclass[12pts,manuscript,a4paper]{elsart}
\usepackage{latexsym}
\usepackage{amssymb}
\usepackage{wasysym}
\usepackage[tight,scriptsize]{subfigure}
\usepackage{natbib}
\usepackage{graphicx}

\newcommand{\aap}{A\&A}
\newcommand{\aj}{AJ}

\newcommand{\apj}{ApJ}
\newcommand{\apjl}{ApJL}

\newcommand{\pasp}{PASP}

\begin{document}

\begin{frontmatter}
\title{UV habitable zones around M stars}
\author[iafe]{Andrea P. Buccino}
\ead{abuccino@iafe.uba.ar}
\author[cea]{Guillermo A. Lemarchand}
\ead{lemar@correo.uba.ar}
\author{Pablo J. D. Mauas}
\ead{pablo@iafe.uba.ar}

\address[iafe]{ Instituto de Astronom\'\i a y F\'\i sica del Espacio (CONICET),
    C.C. 67 Sucursal 28, C1428EHA-Buenos Aires Argentina}
\address[cea]{Facultad de Ciencias Exactas y Naturales, Universidad
    de Buenos  Aires,
    C.C. 8 - Sucursal 25, C1425FFJ Buenos Aires Argentina and Instituto
    Argentino de Radioastronom\'\i a (CONICET), C.C. 5, 1894, Villa Elisa, Buenos Aires, Argentina}

\begin{abstract}
During the last decade there was a change in paradigm, which led to
  consider that terrestrial-type planets within liquid-water habitable
  zones (LW-HZ) around M stars can also be suitable places for the
  emergence and evolution of life. Since many dMe stars emit large
  amount of UV radiation during flares, in this work we analyze the
  UV constrains for living systems on Earth-like planets around dM
  stars. We apply our model of UV habitable zone (UV-HZ, Buccino \emph{et
  al.} 2006) to the three planetary systems around dM stars (HIP 74995,
  HIP 109388 and HIP 113020) observed by IUE and to two M-flare stars
  (AD Leo and EV Lac). In particular, HIP 74995 hosts a terrestrial
  planet in the LW-HZ, which is the exoplanet that most resembles our
  own Earth. We show, in general, that during the quiescent state
  there would not be enough UV radiation within the LW-HZ to trigger the
  biogenic processes and that this energy could be provided by flares
  of moderate intensity, while strong flares do not necessarily
  rule-out the possibility of life-bearing planets.
\end{abstract}
\end{frontmatter}

\section{Introduction}\label{intro}

Dwarfs of spectral type M (dM stars) constitute 75\% of main sequence
stars and, even having relative low masses, they contribute more than
any other spectral type to the total stellar mass of the galaxy
\citep{1986mts..book..409R}.  M stars are much smaller in mass than
the Sun (between 0.08 and 0.5 $M_{\odot}$) and their hydrogen burning
lifetimes are much longer. Their lifetimes range from 50 Gyr to
several trillion years \citep{1997ApJ...482..420L} with a
very slow change in their emitted flux. They also have relatively low
stellar temperatures $(2400 K \lesssim T_{eff} \lesssim 3900
K)$. In the visual range, a typical M0 V spectrum represents
only 1.9 \% of the solar flux, while at the infrared band,
the relation is 16 \%. On the other extreme (e.g. M9 V) the values
are respectively $1.4 \times 10^{-14}$\% and 0.13\%. Therefore,  the balance
of radiation of M stars is very different from our Sun.

Using numerical simulations, \cite{1996Icar..119..219W} showed that
planets are likely to form in the habitable zones (HZ) of M dwarfs. In
the last years, several new proposals to enhance the detection of
planets around M dwarfs were promoted (\citealt{2003AJ....126.3099E},
\citealt{2004sshp.conf..231D}). However, recent surveys have only
detected nine planetary systems around M dwarfs \citep{Sch2006}.  In
particular, \cite{2007arXiv0704.3841U} detected two super-Earth
planets around the HIP 74995 system. One of them (Gl581c) is of great
interest since it has a mass of 5.1 M$_\oplus $ (the lowest mass found
for an exoplanet to date) and resides in the liquid-water habitable
zone (LW-HZ). \cite{2007arXiv0704.3841U} affirm that
Gl581c is the exoplanet yet discovered that most resembles the
Earth. However, \cite{2007arXiv0705.3758V} calculated the habitable
zone constrained by the levels of biological productivity on Gl581c's
surface and found this planet too close to the parent star to host
life.

Habitable planets around the low-mass M stars would have some
significant differences to Earth. In particular, most planets within
the liquid water habitable zones defined in \cite{1993Icar..101..108K}
are probably tidally locked, with the same side always facing the
central star (\citealt{1964QB54.D63}, \citealt{1993Icar..101..108K},
\citealt{2003AsBio...3..415J}). Early arguments assumed that
atmospheric volatiles would freeze in the dark side and boil off on
the side of the planet facing the star. In this way, the the
atmospheric pressure at the surfacewould be orders of magnitude below
present-dayEarth and far below the minimum pressure at which liquid
water can exist. However, according to different models of increasing
complexity developed by \cite{1996chz..conf...29H},
\cite{1997Icar..129..450J}, \cite{1999OLEB...29..405H} and
\cite{2003AsBio...3..415J}, atmospheric heat transport could prevent
freezing on the dark side of the planet. Therefore, it could be
possibleto have habitable synchronously rotating planets where liquid
water can exist.

Another important characteristic is that starspots may induce
rotationally-modulated variations of several percent in the M-stellar
radiation on time-scales of days (\citealt{1986mts..book..409R},
\citealt{1998AJ....116..429B},
\citealt{2007arXiv0707.2577K}). Although large starspots may cause a
significant decrease in the stellar brightness, an atmosphere with a
pressure of 1 bar at the surface of the planet would not freeze
\citep{1997Icar..129..450J}.  Atmospheric collapse would occur only
if the planet were at the coldest CO$_2$ condensation end of the
habitable zone \citep{1993Icar..101..108K}.

In a recent review, \cite{2006astro.ph..9799T} concluded that M dwarf
stars  may indeed be viable hosts for planets on which the origin and
evolution of life can occur. They have also found a number of
planetary  processes such as cessation of geothermal activity, or
thermal and  non-thermal atmospheric loss processes that may limit the
duration of planetary habitability to times far shorter than the
lifetime of  the M dwarf star.

\cite{Scaloetal} describe several sources of environmental
fluctuations generated by dMe flares with strong influence on their
habitability conditions. These include short-term variations in UV and
blue radiation at the surface, effects due to alterations of the
atmospheric photochemistry, heating of the atmosphere and surface by
the effects of strong flares and energetic particles that could have
intermittent effects on ozone and other chemistry. 

It is well known that in the Sun there is a close correlation
between strong flares and coronal mass ejections
(CME). \cite{2007AsBio...7..167K} analyzed the impact of stellar CMEs
on Eath-like planets around active M stars. They found that those
exoplanets within the habitable zones close ($<0.01$ AU) to the dMe
star would be exposed to the effects of CMEs for a long period. Since
most of these exoplanets are tidally locked, their intrinsic magnetic
moments are weak and the magnetosphere can be compressed under the
action of CME plasma flow. This interaction would cause strong
atmospheric erosion on Earth-like exoplanets.

In particular, UV radiation between 200 and 300 nm is known to inhibit
photosynthesis, induce DNA destruction and cause damage to a wide
variety of proteins and lipids (\citealt{LH91},
\citealt{Cock1998}). On the other hand, UV radiation is thought to
have played an important role in the origin of life
({\citealt{Toup77}, \citealt{2002RPPh...65.1427E}). It is usually
considered one of the most important energy sources on the primitive
Earth for the synthesis of many biochemical compounds. Based on these
considerations and the \emph{Principle of Mediocrity}\footnote{The
so-called \emph{Principle of Mediocrity} proposes that our planetary
system, life on Earth and our technological civilization are about
average and that life and intelligence will develop by the same rules
of natural selection wherever the proper conditions and the needed
time are given \citep{1961Sci...134.1839V}.}, in
\cite{2006Icar..183..491B} we defined the boundaries of an ultraviolet
habitable zone (UV-HZ).  In that work, we also analyzed the UV-HZ
on solar-type stars with exoplanets observed by the
\emph{International Ultraviolet Explorer} (IUE) satellite.

In the present work, we extend our analysis by considering the
UV constrains for the emergence and evolution of living systems for
all the M stars with exoplanets that have been observed by IUE.We
compare these results with the LW-HZ. Since many dM flare stars emit
large amounts of ultraviolet radiation and X-rays during flares, we
also analyze how the UV Habitable Zones behave with the presence of
moderate and strong flares.

 In \S \ref{obs} we describe the methodologies used to
perform the observations and data analysis. In \S \ref{HZs} we
apply our model to the dM planetary systems and to two dM flare stars and
in \S \ref{discuss} we present a discussion of the results.

\section{Observations}\label{obs}

To date (June 2007) nine dM stars were found hosting planetary
systems.  Three of them were observed by IUE: HIP 74995 (M3V), HIP
109388 (M3.5V) and HIP 113020 (M4V). In Table \ref{planet_data} we
list some physical parameters of the host stars and their
corresponding planets.  The flare stars AD Leo (M3.5Ve) and EV Lac
(M4.5Ve) were also studied to analyze the influence of flares on
living systems.

\textbf{[Table \ref{planet_data}]}

To study the influence of near UV radiation to exoplanets around M
stars, we use IUE low-(0.6 nm resolution) and high-dispersion
($\lambda/\Delta\lambda\sim$10000) spectra, taken by the long
wavelength cameras (LWP and LWR) in the range 185-340 nm. The spectra
are available inthe IUE public library (at
http:\emph{//ines.laeff.esa.es/cgi-ines/IUEdbsMY}), and have been
calibrated using the NEWSIPS (New Spectral Image Processing System)
algorithm. The internal accuracy of the high-resolution calibration is
around 4\% (\citealt{2000A&AS..141..331C}) and the errors of the low-
dispersion spectra in the absolute calibration are around 10-15\%
\citep{1998AAS...193.1122M}.

For each of the three dM stars of our sample, only one IUE low-resolution
spectra is available. On the other hand, for EV Lac there are 54 IUE
low-resolution spectra in the long-wavelength range, and for AD Leo 64
low- and high-resolution spectra in the same wavelength range. 

\textbf{[Figure \ref{fig.IUE_esp}]}

To illustrate the dependence of UV stellar radiation on its spectral
class, in Fig. \ref{fig.IUE_esp} we show a selection of IUE
spectra scaled at 1 AU corresponding to the following stars of
different spectral types: HD 9826 (F8V), HD 3651 (K0V) and the solar
twin 18 Scorpii (G2V), together with the stars under study here. We
also include two IUE spectra of the very active star AD Leo (M3.5Ve)
in the quiescent state and at the maximum of a strong flare.  It is
worth mentioning that HIP 113020 (M4V) was reported as an inactive M
star by \cite{JohnsKrull96}, HIP 74995 (M3V) presents
signatures of weak chromospheric activity \citep{2005A&A...443L..15B}
and HIP 109388 (M3.5V) was reported as a middle-aged dwarf star of low
activity \citep{2006PASP..118.1685B}.  Therefore, the dM spectra
in Fig \ref{fig.IUE_esp} show different levels of stellar
activity.

We can see in Fig. \ref{fig.IUE_esp} that the level of UV radiation
is much  lower in dM stars than in  G stars.
In particular, due to their low effective temperature, dM spectra are
dominated by molecular absorption bands.



\section{Habitable zones around M stars}\label{HZs}

We computed the LW-HZ (\citealt{1993Icar..101..108K}) and the
UV-HZ as described in \cite{2006Icar..183..491B} for all the M star
with exoplanets observed by IUE. We refer the reader to the
latter paper for details. In what follows we summarize the
expressions we used in \cite{2006Icar..183..491B}.

As mentioned in \S \ref{intro}, it is believed that UV radiation
 played an important role in the  Earth biogenic processes
 (\citealt{Toup77}, \citealt{2002RPPh...65.1427E}). Consequently,
 there should be a minimum number of UV photons for the
 chemical synthesis of complex molecules to happen also in
 exoplanets. Based on the \emph{Principle of Mediocrity}, we set the
 outer limit of the UV-HZ ($d_{out}$), imposing that:
\begin{equation}
N^{*}_{UV}(d_{out}) \ge 0.5 \times N^{\odot}_{UV}(1
\textrm{\footnotesize{AU}\normalsize})|_{t=t^{\odot}_{Arc}}.\label{outlim}
\end{equation}
where $N^*_{UV}(d_{out})$ is the UV flux  at a distance
$d_{out}$ photons emitted by the star in the wavelength 200-315 nm,
and $ N^{\odot}_{UV}(1
\textrm{\footnotesize{AU}\normalsize})|_{t=t^{\odot}_{Arc}}$ is the
flux of UV photons received on top of the atmosphere of Primitive
Earth when life emerged without any ozone layer protection.

On other the hand, UV radiation could be damaging for biological
systems.  The destructive effect of the UV radiation on biochemicals
processes is usually considered through a biological action spectrum
(BAS) B($\lambda$), which represents a relative measure of damage as a
function of wavelength. In \cite{2006Icar..183..491B}, we have defined
B($\lambda$) as a function proportional to the probability of a photon
of energy $\frac{hc}{\lambda}$ to dissociate free DNA, given by the
following expression:
\begin{equation}
\centering
log\,B(\lambda)\sim\frac{6.113}{1+\exp(\frac{\lambda[nm]-310.9}{8.8})}\,-4.6.
\end{equation}

Again, applying  the \emph{Principle of Mediocrity}, we set the
inner limit $d_{in}$ of the UV-HZ imposing that:
\begin{equation}
\centering N^{*}_{DNA}(d_{in}) \le 2 \times N^{\odot}_{DNA}(1
\textrm{\footnotesize{AU}\normalsize})|_{t=t^{\odot}_{Arc}},\label{innlim}
\end{equation}
where $N^{*}_{DNA}(d_{in})$ is the flux of damaging photons at
a distance $d_{in}$, weighted with $B(\lambda)$, and
$N^{\odot}_{DNA}(1
\textrm{\footnotesize{AU}\normalsize})|_{t=t^{\odot}_{Arc}}$ is the
flux of DNA damaging photons received on Earth 3.8 Gyrs ago.
All the atmospheric UV attenuation can be neglected compared to the
factor of 2 used due to the \emph{Principle of Mediocrity}
\citep{2006Icar..183..491B}.

In Fig. \ref{fig.HZ_M} we plot the UV-HZ for the stars under study. We
also show the LW-HZ. We note that the LW-HZ around M stars can be
wider than the one considered by \cite{1993Icar..101..108K}. For
example, \cite{2003AsBio...3..415J} reported that the planetary albedo
of an ocean-covered synchronously- rotating Earth would be
20\% higher than if the planet is not tidally locked. This higher
albedo could make the inner limit of the LW-HZ a 10\% closer to the
star than the traditional value.  On the other hand, the UV
radiation of M stars could lead to a different photochemistry in the
planetary atmosphere (richer in CH$_4$ and N$_2$O than Earth) and
could move away the outer edge of the LW-HZ
\citep{2005AsBio...5..706S}. However, this phenomena is not
quantified.  Therefore, in our LW-HZ calculations we adopt
the less restrictive boundaries criteria considered by
\cite{1993Icar..101..108K}, which should be a good approximation for
dM stars (for our solar system are 0.75 and 1.77 AU) .

\textbf{[Figure \ref{fig.HZ_M}]}

\cite{1993ApJ...403..303L} presented several evolutionary tracks which
show that the luminosity of stars with masses 0.25-0.35
$M_{\odot}$ is almost constant during the main sequence stage. For
this reason, we do not need to simulate the temporal evolution of the
habitable zones. Both the LW-HZ and the UV-HZ plotted in
Fig. \ref{fig.HZ_M} would remain constant for near 100 Gyr
(\citealt{1993Icar..101..108K},
\citealt{2006astro.ph..9799T}). Low-amplitude short-scale
variations, no longer than hours, may arise from flare activity,
whereas starspots may induce variations of several percent on
time-scales from days (rotational modulation of spots) to years
(starspots cycle). 

For the giant planet Gl876c, the presence of liquid water on a
hypothetical moon is possible.  Similarly, the exoplanet Gl581c,
which is a \emph{super-Earth} with $\sim 5 M_{\oplus}$, resides in the
LW-HZ of the star. As pointed in \S\ref{intro}, this is an important
case as it is the planet which most resembles our own Earth.  However,
neither of these two planets receive enough UV radiation to start the
biogenic processes, an alternative energy source would be needed to
trigger the formation of complex molecules for the origin of
life. This conclusion also applies to any hypothetical terrestrial
exoplanet in the LW-HZ around HIP 109388.

On the other hand, much larger UV fluxes can reach the planetary
atmosphere during stellar flares, which for dMe stars can be very
intense. To study the biological influence of stellar flares, we
included in Fig. \ref{fig.HZ_M} the UV-HZ and the LW-HZ of two well
studied flare stars (AD Leo, M3.5Ve, and EV Lac, M4.5Ve). In both
cases, we have estimated the UV-HZ in the quiescent state and for
flares of different strength.  During the strong flare in AD Leo the
UV-HZ increased its width by a factor $\sim$6, while the inner boundary moved
to a position 5.5 times farther than in the quiescent state. In the
case of the weaker flare in EV Lac, those factor were 2 and 2.2
respectively.  The biological implications of this fact will be
discussed in the next Section.

\section{Discussion}\label{discuss}

In the three exoplanetary cases around M stars observed by IUE, the
LW-HZ and the UV-HZ are completely separated.  Since these
stars have low levels of activity, the UV-HZ would not present
significant variations in time.  The UV radiation of these dM stars
within the LW-HZ is orders of magnitude smaller than that required to
trigger the formation of complex molecules.

In recent years, a strong debate took place in the
astrobiological literature about the possibility of panspermia or
migration of life-seeds among nearby planets. The distances among dM
exoplanets could be shorter than planets around other stellar types,
 a fact which could increase the chances of panspermia. In this
way, life could originate in a planet with enough UV radiation and
eventually migrate to a planet with less UV radiation but more
moderate temperatures. This could be the case of Gl581 and Gl876
planetary systems.

Nevertheless, for the origin of life on an
exoplanet in the LW-HZ, a different energy source or an alternative
physical mechanism is needed. 
In principle, flares could provide the energy for the
biogenesis. There are large variations regarding the duration,
frequency and energy released during stellar flares. In all cases,
however, the most energetic impulsive phase of the flare lasts from
fractions of a second to a few minutes, and the decay phase, much less
energetic, lasts from seconds to many hours. Regarding the flare
frequency, it is well known that large flares occur less often and
usually last longer than smaller ones (see
e.g. \citealt{2005stam.book.....G}).

As an example of a \emph{moderate} flare, we consider the one that
took place on EV Lac on September 10$^{\textrm{th}}$ 1993
\citep{1995AAS...186.2103P}. It released a UV flux (in the center of
the LW-HZ) of F$_{UV}$(200-315 nm)=4.94$\times$10$^3$erg cm$^2$
s$^{-1}$.  For a \emph{strong} flare, we considered the one observed
in AD Leo on April 12$^{\textrm{th}}$ 1985, which lasted more than 4
hours, with an abrupt brightness increase and a long decay
(\citealt{1991ApJ...378..725H}, \citealt{1996A&A...310..245M}).  At
the peak of this flare, a UV flux F$_{UV}$(200-315
nm)=4.51$\times$10$^4$erg cm$^2$ s$^{-1}$ reached the center of the
LW-HZ.

In Fig. \ref{fig.HZ_M} we show the impact of these flares on the
UV-HZ.  It can be seen that during the moderate flare in EV Lac, the
UV-HZ  coincides with the LW-HZ. Therefore, a moderate flare could
provide the UV energy necessary to trigger the biogenic
processes. However, since flares are sporadic and of short duration,
this energy would be available for only a small fraction of the
time. Therefore, it should be much less probable to originate life in a
terrestrial planet in the LW-HZ of dM stars.

\textbf{[Figure \ref{fig.ADL-time}]}

On the other hand, the UV radiation emitted by the strong flare in AD
Leo is several orders of magnitude larger than the radiation needed to
destroy biomolecules and, therefore, could seriously affect the
development of living systems.

However, flares only last a few  hours, and
most part of this time is spent on its long decay, when the energy
emitted is much lower. To illustrate this point, in
Fig. \ref{fig.ADL-time} we  show the UV
light curve (at 1 AU) for this particular flare on AD Leo. It can be
seen that the UV flux is already reduced by a factor of 10 about 1 hour after peak
time. Also shown in Fig. \ref{fig.ADL-time} is the flux of
DNA-damaging UV  photons $N^*_{DNA}$, which evolves with a similar pattern.
Moreover, most planets within the LW-HZ around dM stars are
probably tidally locked and, therefore, only one face of the planet would
receive the damaging UV radiation. Strong flares could  provide a
strong selective pressure for the emergence and evolution of living
organisms, but not necessarily preclude their existence.

On the other hand, it should be kept in mind that strong flares are
much less frequent than moderates ones. In particular, the strong AD
Leo flare studied here was one of the strongest flares ever recorded
\citep{1991ApJ...378..725H}. Therefore, mechanisms of DNA repair could
operate in the time between damaging flares. In consequence, the
intermittent variation of UV radiation emitted by flares could be a
source of higher mutation rates. This, in turn, would imply higher
microbial biodiversity, and faster adaptation to the changing UV
environmental conditions.

\cite{2005AsBio...5..706S} made several simulations of the evolution
 of atmospheres of hypothetical planets around chromospherically
 active M stars, and found that the relatively large UV flux could be
 greatly attenuated by ozone layers similar to the terrestrial one and
 even thicker (see Table 1 in \citealt{Scaloetal}). However, their
 work refers to atmospheric concentrations similar to the present
 terrestrial one (1 PAL of O$_2$ and N$_2$), while life is supposed to
 have origined with much lower oxygen concentrations.

Another factor to consider is the evolution of chromospheric
activity of M-type stars. The usually accepted model to describe the
generation and intensification of magnetic fields in late F- to early
M-type stars is the $\alpha\Omega$-dynamo first invoked to explain
solar activity (\citealt{1955ApJ...122..293P}). This model,
where the large-scale magnetic field generation results from the
interaction of differential rotation in the tachocline and the
convective turbulence,  predicts a strong correlation between
activity and rotation.  Magnetic activity, which is closely related to
stellar rotation, decays with time as the star spins down due to
braking by magnetized winds. However, recent results show that this
decay happens in the first two Gyr of the stellar life, to be compared
with lifetimes for these stars which are at least 25 times larger
\citep{2004A&A...426.1021P}.

This model can only explain the
magnetic behaviour of F to early M stars, since stars later than M3
are believed to be fully convective and, therefore, cannot sustain a
solar like $\alpha\Omega$ dynamo. In particular, the three M stars
with exoplanets (HIP 74995, HIP 109388 and HIP 113020) are at or
beyond this limit.  Nevertheless, there is plenty of observational
evidence that slow late-type rotators like dMe stars are very active
and have strong magnetic fields, with filling factors larger than for
earlier stars (\citealt{1980ApJ...239L..27M},
\citealt{1989PhDT.........1H},
\citealt{1996A&A...310..245M}). \cite{2006A&A...446.1027C} proposed
that for fully convective M stars, large-scale magnetic fields could
be produced by a pure $\alpha^2$ dynamo, where activity would not
decay with time since it does not involve rotation.  Although, this
model does not predict a cyclic activity, recently
\cite{2007A&A...461.1107C} found an activity cycle for Proxima
Centauri (M5Ve) with a period of $\sim$1.2 year.  In any case, Prox
Cen is a slow rotator, 4-4.5 Gyr old \citep{1986ApJ...300..773D}, and
\emph{very active}: it presents a flare every 10 hours
\citep{2007A&A...461.1107C}.

In summary, our results show that terrestrial-type planets within the LW-HZ
around inactive M stars do not receive enough UV radiation to perform
the synthesis of complex macromolecules, and would therefore need an
alternative energy source to start the biogenesis. In particular, this is the
case of Gl581c, which, according to \cite{2007arXiv0704.3841U}, is the
planet most similar to Earth yet discovered and considered, to date,
one of the best candidates for hosting life. In contrast to what it
has been believed for a long time, moderate flare activity could play
an important role in the origin and evolution of life, triggering the
biogenic processes, while the effect of strong flares, that are less
frequent, could be mitigated by the fact that most exoplanets within
the LW-HZ are probably tidally locked, and therefore only one face of
the planet would receive the damaging UV radiation.  Taking everything
into account, therefore, M stars with moderate flares are the best
candidates to host habitable planets.

\textbf{Acknowledgements}

This research was supported by U401 and X271 UBACYT Research Projects
from the University of Buenos Aires and Project 03-12187 of the
Agencia de Promoci\'on Cient\'\i fica y Tecnol\'ogica (Argentina). G.A.L. was
partially supported by \emph{The Planetary Society} SETI Grant. A.B. was
sponsored by a CONICET graduate scholarship.

We thank Sergio Messina and the anonymous Reviewer II for their useful
comments and observations.We also thank Antigona Segura and Jill
Tarter for their useful comments to an early version of this work.


\newpage

\textbf{Tables}
\renewcommand{\baselinestretch}{1.}
\begin{table*}[!htb]
\caption{Physical Parameters of M stars with exoplantes observed by
  IUE and their planets.}\label{planet_data}
\begin{tabular}{r r c c l c l c c r}
\hline
\hline
\multicolumn{6}{l}{Stellar Properties}&
\multicolumn{4}{l}{Planetary Properties} \\
\hline
Star & Sp. Type &Mass &dist. & m$_V$ & Age &  Planet &Mass[sin$i$]&Semimajor & Period\\
HIP     & and Class         & (M$_\odot$) &(pc)  &       & (Gyr.) &  &  (M$_J$) &Axis (AU) & (days)\\
\hline \\
74995 & M3V  & 0.31&6.26 &	10.55	&4.30  &	Gl581b & 0.060  &0.04 &5.40 \\
      &        &    &     &             &      &      Gl581c & 0.016 & 0.07 & 12.93\\
      &        &    &     &             &      &      Gl581d & 0.024 & 0.25 & 83.60\\
109388 &	M3.5V &0.36 &	8.80 & 10.42 &	-	& Gj849b &0.820  &2.35 &1890.00 \\
113020 &M4V &0.32 &4.72 & 10.17& 9.90 &Gl876b  & 1.940 &	0.21 &	60.94\\

      &      &      &   &    &           &	Gl876c &0.560 &	0.13 &	30.10\\
       &   &     & &     &     & Gl876d	&0.020 &0.02  &	1.94 \\

\hline
\end{tabular}
\end{table*}
\renewcommand{\baselinestretch}{2.}
\newpage

\textbf{FIGURES CAPTIONS}

\underline{Caption Fig. \ref{fig.IUE_esp}}: IUE Spectra of F to M
  dwarfs stars: HD 9826 (F8V, short dash), 18 Sco (G2V, solid), HD
  3651 (K0V, long dash), AD Leo quiescent (M3.5V, dot-long dash), AD
  Leo with flare (M3.5Ve, dot-short dash), and, with dotted line, the
  M stars with exoplanets HIP 109388 (M3.5V), HIP 113020 (M4V) and HIP
  74995 (M3V). The last two spectra are arbitrary displaced for
  clarity by $10^{-2}$ and $10^{-4}$ erg cm$^{-2}$ s$^{-1}$\AA$^{-1}$
  respectively.

\underline{Caption Fig. \ref{fig.HZ_M}}: Habitable zones around the
planetary stars HIP 109388 (Gl 581), HIP 74995 (Gj 849) and HIP 113020
(Gl 876) and around the flare stars AD Leo and EV Lac in the quiescent
and flaring states.  The solid lines represent the LW-HZ, the
rectangles are the UV-HZ and the triangular dots indicate the
positions of the exoplanets.

\underline{Caption Fig. \ref{fig.ADL-time}}: The UV flux at 1 AU ($\blacktriangle$) and
  the  corresponding flux of DNA-damaging
UV photons $N^*_{DNA}$ at 1 AU ($\bullet$)  of AD Leo flare of April,
  12$^{\textrm{th}}$  1985.

\newpage
\textbf{FIGURES}

\begin{figure}[htb!]
\centering
\includegraphics[width=0.95\textwidth]{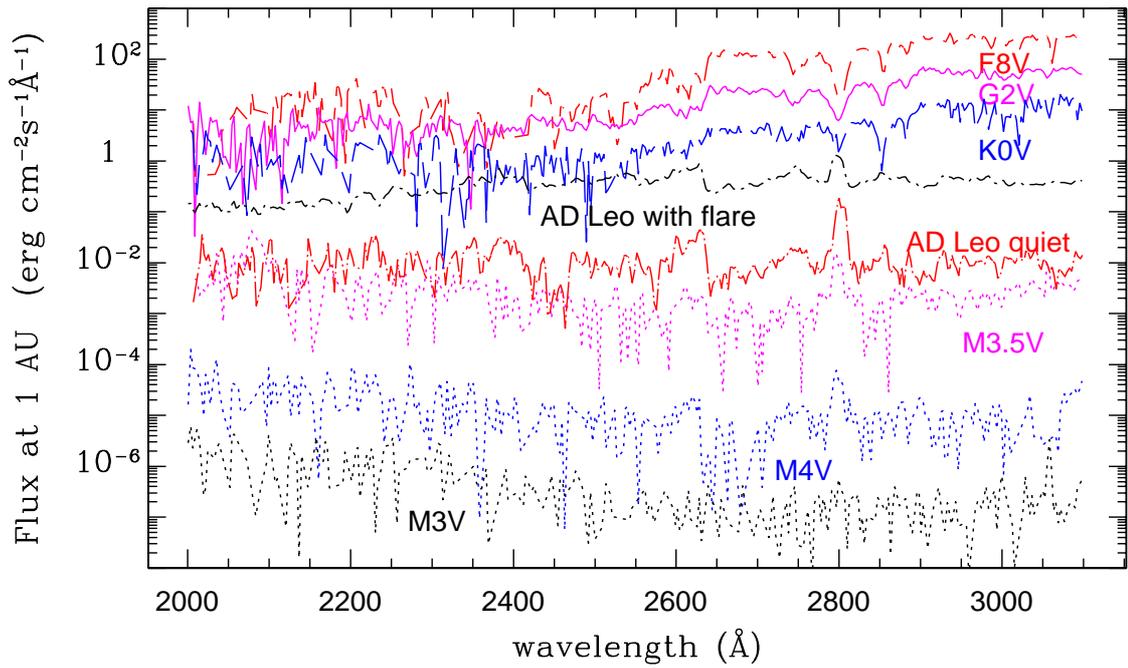}
\caption{Buccino \emph{et al.}, 2007. IUE spectra of M
stars.}\label{fig.IUE_esp}
\end{figure}
\renewcommand{\baselinestretch}{2.}

\begin{figure}[htb!]
\centering
\includegraphics[width=0.75\textwidth]{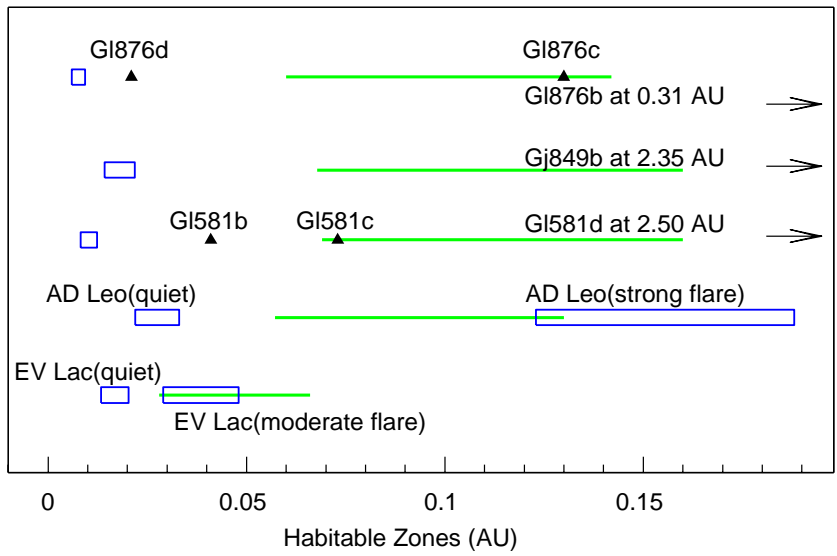}
\caption{Buccino \emph{et al.}, 2007. Habitable Zones around M stars.}\label{fig.HZ_M}
\end{figure} 

\begin{figure}[htb!]
\centering
\includegraphics[width=0.75\textwidth]{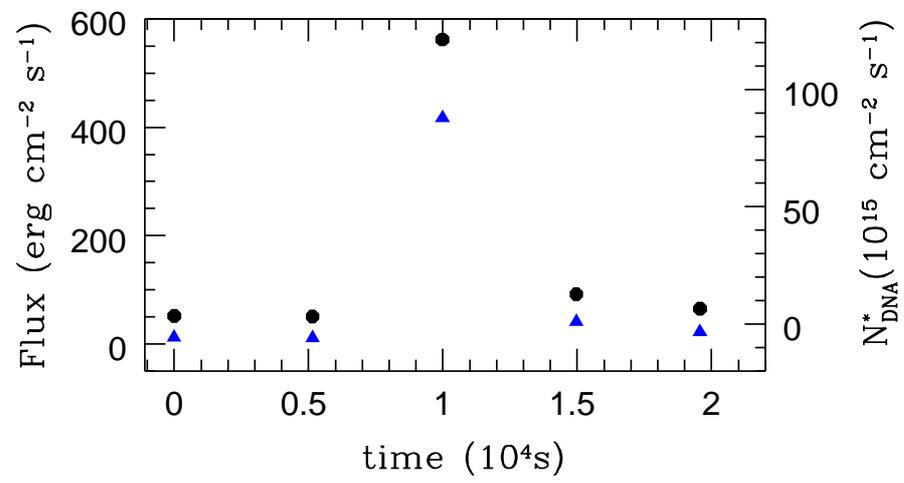}
\caption{Buccino \emph{et al.}, 2007. Flare in AD Leo.}\label{fig.ADL-time}
\end{figure}
\renewcommand{\baselinestretch}{2.} 

\end{document}